\documentclass[
    ,final            
  ]
  {aipproc}

\layoutstyle{8x11double}

\begin{document} 

\title{Boost symmetry in the Quantum Gravity sector} 



\author{Francesco Cianfrani}
{address={ICRA-International Center for Relativistic Astrophysics
\\Dipartimento di Fisica (G9), Universit\`a di Roma, ``Sapienza'',
\\Piazzale Aldo Moro 5, 00185 Rome, Italy.}} 

\author{Giovanni Montani}
{address={ICRA-International Center for Relativistic Astrophysics
\\Dipartimento di Fisica (G9), Universit\`a di Roma, ``Sapienza'',
\\Piazzale Aldo Moro 5, 00185 Rome, Italy.}
,altaddress={ENEA C.R. Frascati (Dipartimento F.P.N.), via Enrico Fermi 45, 00044 Frascati, Rome, Italy.\\
and\\
ICRANET C. C. Pescara, Piazzale della Repubblica, 10, 65100 Pescara, Italy.}}

\begin{abstract}
We perform a canonical quantization of gravity in a second-order formulation, taking as configuration variables those describing a 4-bein, not adapted to the space-time splitting. We outline how, neither if we fix the Lorentz frame before quantizing, nor if we perform no gauge fixing at all, is invariance under boost transformations affected by the quantization.  \end{abstract} 

\maketitle 


\section{Introduction}
The development of a quantum theory for the gravitation field is one of the main task of Theoretical Physics. The standard canonical quantization \cite{DW67} failed on a technical level, since a precise definition of the Hilbert space structure has not been given. Moreover, some issues of the classical formulation have not been solved, as for instance the avoidance of singularities. In the last twenty years a great effort to Quantum Gravity has been given by Loop Quantum Gravity program (for a review see \cite{Th01}). On a kinematical level, the main achievement of this formulation is a natural derivation of a discrete structure for 3-geometries, in terms of discrete spectra for area and volume operators. This point is at the basis of other convincing results in mini-superspace models, as the avoidance of cosmological singularity \cite{Bo}. Furthermore, on theoretical ground, a discrete space-time is an expected feature. But a clear space-time characterization for this discreteness has not been given. In particular, since Loop Quantum Gravity is based on a formulation which employs a fixing of the boost symmetry, it is not clear if discrete spectra are modified by boosts.\\  
In this work we are starting an investigation towards the behavior of geometrical operators spectra, by performing a second order formulation for gravity, with a formal gauge fixing. This way quantum transformation between Hilbert spaces, corresponding to different sectors, can be investigated and it will be probed that boosts can be realized on a quantum level as a symmetry.\\ 
The paper is organized as follows: in section 1 the 3+1 splitting of the space-time, by which a time parameter arises, is presented, so that in section 2 the standard Hamiltonian formulation, {\it i.e.} a second order formulation in metric-like variables, is provided. Then section 3 deals with the Holst reformulation of General Relativity, which is at the basis of Loop Quantum Gravity, and in section 4 the implications of this model in view of the boost invariance are discussed. In section 5 we develop a formulation suitable for our aim, such that vier-bein variables can be introduced without any restriction. The properties of such dynamical system are investigated in section 6, while in section 7 the fixing of the Lorentz frame is performed and the theory is canonically quantized. In section 8 quantum boosts, relating different sectors, will be developed and they will be realized as symmetry transformations. In section 9 brief concluding remarks follow.

\section{The space-time slicing} 
In order to provide an Hamiltonian formulation for the gravitational field, a time parameter has to be identified. We are going to describe the standard procedure, by which a formal splitting of the space-time is performed, {\it i.e.} the Arnowitt-Deser-Misner (ADM) formulation \cite{ADM59,ADM60a}.\\
In order to have a well-posed initial value formulation for the gravitational field, the basic assumption is to deal with a global hyperbolic space-time $V^4$. This request means that a Cauchy hypersurface exists, {\it i.e.} there is a surface such that the evolution backwards and forward of initial conditions on it are enough to provide the form of the metric on the full manifold. Within this scheme, a diffeomorphism $y^\mu=y^\mu(t, x^i)\hspace{0.2cm}(i=1,2,3)$ can be defined mapping $V^4$ into $\Sigma\otimes R$ \cite{G70}, $\Sigma$ being a space-like hypersurfaces with coordinates $x^i$, while $t$ is the time-like real one. This diffeomorphism provides the space-time splitting.\\ 
Given the normal vector to $\Sigma$, $\eta$, and the tangent ones $\vec{e}_i=\frac{\partial y^\mu}{\partial x^i}\partial_\mu$, the following relations come from the space-like character of $\Sigma$ 
\begin{eqnarray}
\left\{\begin{array}{c} \vec{\eta}\cdot\vec{\eta}=-1\\ \vec{\eta}\cdot\vec{e}_i=0 \\ \vec{e}_i\cdot\vec{e}_j=h_{ij} 
\end{array}\right.\Rightarrow 
\left\{\begin{array}{c} \eta^\mu\eta^\nu g_{\mu\nu}=-1\\ \eta^\mu e^\nu_i g_{\mu\nu}=0 \\ e^\mu_i e^\nu_j g_{\mu\nu}=h_{ij} \end{array}\right.
\end{eqnarray} 
$h$ being a positive-definite symmetric matrix. As soon as the deformation vector $\vec{e}_0=\partial_t=\partial_t y^\mu\partial_\mu$ is introduced, the lapse function $\widetilde{N}$ and the shift vector $\widetilde{N}^i$ can be defined as 
\begin{eqnarray}
\vec{e}_0=\frac{\partial u^\mu}{\partial x^0}\vec{f}_\mu=\widetilde{N}\vec{\eta}+\widetilde{N}^i\vec{e}_i,\label{e0}
\end{eqnarray}
$\widetilde{N}$ being non-vanishing.\\
The 3+1 splitting of the metric tensor can now be performed, obtaining
\begin{eqnarray}
\left\{\begin{array}{c} g_{00}=\vec{e}_0\cdot\vec{e}_0=-\widetilde{N}^2+h_{ij}\widetilde{N}^i\widetilde{N}^j\\ g_{0i}=\vec{e}_0\cdot\vec{e}_i=h_{ij}\widetilde{N}^j \\ g_{ij}=\vec{e}_i\cdot\vec{e}_j=h_{ij} \end{array}\right..\label{metrsplit}
\end{eqnarray}
The next step is the projection on each $\Sigma$ of all 4-quantities. In this respect, an object $q^{\mu\nu}=h^{ij}e_i^\mu e_j^\nu$ can be defined and, starting from the completeness relation for the basis $(\eta,e_i)$, {\it i.e.} \begin{equation}
-\eta^\mu\eta^\nu+h^{ij}e_i^\mu e_j^\nu=g^{\mu\nu},\label{comp}
\end{equation}
one recognizes it as a projector, since $q^{\mu\nu}\eta_\mu=0$.\\
The covariant derivative on $\Sigma$ can also be splitted; it is defined according with the formula\begin{equation}\vec{e}_i\cdot(\partial_j\vec{A})=D_jA_i=\partial_jA_i+\Gamma^k_{ij}A_k,\label{De}\end{equation}$A_k$ being a spatial vector, while $\Gamma^k_{ij}$ are 3-dim affine connections, whose relation with 4-dimensional ones can be obtained from the geometrical definition
\begin{equation}
\partial_j\vec{A}=(\partial_jA^i)\vec{e}_i+A^i\partial_j\vec{e}_i=(\partial_jA^i)\vec{e}_i+A^i\Gamma^k_{ij}\vec{e}_k+\Pi_{j}\vec{\eta}.
\end{equation}
In fact, in terms of space-time indexes, $D_jA_i$ can be rewritten as
\begin{equation}
D_\mu A_\nu=\frac{\partial x^i}{\partial u^\mu}\frac{\partial x^j}{\partial u^\nu}D_jA_i=e^i_\mu e^j_\nu D_jA_i,
\end{equation}
from which and from the relation (\ref{De}), the following condition comes out
\begin{equation}
D_\mu A_\nu=q^\rho_\mu q^\sigma_\nu \nabla_\rho A_\sigma.\label{3D}
\end{equation}
Therefore, the 3-dimensional covariant derivative (\ref{De}) is the projection by $q^{\mu\nu}$ of the 4-dim covariant one.\\ The calculation of the Hamiltonian involves the extrinsic curvature $K_{\mu\nu}$, {\it i.e.}
\begin{equation}
K_{\mu\nu}=q^\rho_\mu q^\sigma_\nu \nabla_\rho \eta_\sigma\label{excurv}
\end{equation}
whose projections on $\Sigma$ reads as
\begin{equation}
K_{ij}=(\partial_i\vec{\eta})\cdot\vec{e}_j.
\end{equation}
A more useful expression of $K_{ij}$, by virtue of definition (\ref{e0}), is the following one
\begin{equation} 
K_{ij}=\frac{1}{2\widetilde{N}}[-\partial_0h_{ij}-D_i\widetilde{N}_j-D_j\widetilde{N}_i]\label{excurv1}
\end{equation}
which clarifies that it is symmetric under changes of indexes.\\ 
From a geometrical perspective, $K_{ij}$ expresses the curvature of $\Sigma$, seen in a 4-dimensional point of view, since 
\begin{equation}
\partial_i \vec{A}=(D_{i}A^k)\vec{e}_k-K_{ki}A^k\vec{\eta}.
\end{equation}
The 4-dimensional Riemann tensor can now be written in terms of 3-dimensional quantities; in particular, the 3-dimensional Riemann tensor is defined by 
\begin{equation}
{}^{(3)}\!R^\sigma_{\phantom1\rho\mu\nu}A_{\sigma}=[D_\mu;D_\nu]A_\rho,
\end{equation}
which reads, from relations (\ref{comp}) and (\ref{3D}), as follows
\begin{eqnarray}
D_\mu D_\nu A_\rho=q^\alpha_\mu q^\beta_\nu q^\gamma_\rho\nabla_\alpha(q^{\beta'}_\beta q^{\gamma'}_\gamma\nabla_{\beta'}A_{\gamma'})=q^\alpha_\mu q^\beta_\nu\nonumber\\q^\gamma_\rho(q^{\beta'}_\beta q^{\gamma'}_\gamma\nabla_\alpha\nabla_{\beta'}A_{\gamma'}-
q^{\beta'}_\beta\nabla_\alpha(\eta_\gamma\eta^{\gamma'})\nonumber\\
\nabla_{\beta'}A_{\gamma'}-q^{\gamma'}_\gamma\nabla_\alpha(\eta_\beta\eta^{\beta'})\nabla_{\beta'}A_{\gamma'}).
\end{eqnarray}
As far as the extrinsic curvature definition (\ref{excurv}) is taken into account, one gets
\begin{equation}
{}^{(3)}\!R^\sigma_{\phantom1\rho\mu\nu}A_{\sigma}=q^\alpha_\mu q^\beta_\nu q^\gamma_\rho q_\lambda^\delta R_{\delta\gamma\alpha\beta}A^{\lambda}+(K_{\mu\lambda}K_{\nu\rho}-K_{\nu\lambda}K_{\mu\rho})A^\lambda,
\end{equation}
and, finally, the expression for the 3-dimensional scalar curvature reads 
\begin{equation}
{}^{(3)}\!R=q^{\mu\nu}q^{\lambda\rho}R_{\nu\rho\mu\lambda}-(K^2-K_{\mu\nu}K^{\mu\nu}),\label{3R}
\end{equation}
$K$ being the trace of $K_{\mu\nu}$. The first term on the right side can be rewritten according with formulas
\begin{equation}
R=q^{\mu\nu}q^{\lambda\rho}R_{\nu\rho\mu\lambda}+2q^{\mu\nu}\eta^\lambda\eta^\rho R_{\nu\rho\mu\lambda}\label{R}
\end{equation}
and
\begin{eqnarray}
2q^{\mu\nu}\eta^\lambda\eta^\rho R_{\nu\rho\mu\lambda}=2\eta^\mu[\nabla_\nu;\nabla_\mu]\eta^\nu=\nonumber\\=2(K^2-K_{\mu\nu}K^{\mu\nu})+2\nabla_\nu(\eta^\mu\partial_\mu\eta^\nu-\eta^\nu K),\label{pregc}
\end{eqnarray}
so that, because of the expressions (\ref{3R}), (\ref{R}) and (\ref{pregc}), the curvature of the full space-time manifold is expressed in terms of ${}^{(3)}\!R$ as follows (Gauss-Codacci equation) 
\begin{equation} 
R={}^{(3)}\!R+(K^2-K_{\mu\nu}K^{\mu\nu})+2\nabla_\nu(\eta^\mu\partial_\mu\eta^\nu-\eta^\nu K).
\end{equation}
The last term in the relation above is a divergence, hence it does not enter the dynamical description for suitable boundary conditions.\\ 

\section{The Hamiltonian structure}\label{hamstruct}
The Hamiltonian can now be calculated, taking the parameter $t$, labeling spatial hypersurfaces, as time. From (\ref{metrsplit}), the square root of the metric determinant is $\sqrt{-g}=\widetilde{N}\sqrt{h}$, $h$ being the determinant of $h_{ij}$, so that the full action can be written as follows
\begin{equation}
S=-\frac{c^4}{16\pi G}\int dtd^3x \widetilde{N}\sqrt{h}(K^2-K_{ij}K^{ij}+{}^{(3)}\!R),\label{azsplit}
\end{equation}
where surface terms have been neglected, even though their treatment can be highly non-trivial. Configuration variables are the lapse function, the shift vector and the 3-dimensional metric, while their conjugated momenta turn out to be
\begin{equation}
\pi=0\qquad\pi^i=0\qquad\pi^{ij}=\frac{c^4}{16\pi G}\sqrt{h}(Kh^{ij}-K^{ij}), 
\end{equation} 
respectively. The first two conditions behave as primary constraints.\\ 
According with the standard procedure for constrained systems and as soon as Lagrange multipliers $\lambda$ and $\lambda_i$ are considered, the Hamiltonian takes the following form 
\begin{equation}
\mathcal{H}=\int d^3x [\lambda\pi+\lambda_i\pi^i+\widetilde{N}H+\widetilde{N}_iH^i],
\end{equation}
$H^i$ and $H$ being the super-momentum and the super-Hamiltonian, respectively, whose expressions read
\begin{eqnarray}
H^i=-2D_j\pi^{ij}\label{supmom}
\\H=\frac{16\pi G}{2c^4\sqrt{h}}G_{ijkl}\pi^{ij}\pi^{kl}+\sqrt{h}{}^{(3)}\!R,\label{supham}
\end{eqnarray}
with the Super-metric $G_{ijkl}$ equal to $h_{ik}h_{jl}+h_{il}h_{jk}-h_{ij}h_{lk}$.\\ 
By virtue of the canonical symplectic structure, the only non-vanishing Poisson brackets between variables are:
\begin{eqnarray}
\{\pi^{ij}(x^0;x);h_{lm}(x^0;y)\}=\delta^i_{[l}\delta^j_{m]}\delta^3(x-y)\\\{\pi(x^0;x);\widetilde{N}(x^0;y)\}=\delta^3(x-y)\\\{\pi^{i}(x^0;x);\widetilde{N}_j(x^0;y)\}=\delta^i_{j}\delta^3(x-y).
\end{eqnarray}
Given some test-functions $f$ and $f^i$, smeared quantities $\Pi(f)=\int_\Sigma d^3x f\pi$ and $\vec{\Pi}(\vec{f})=\int_\Sigma d^3x f_i\pi^i$ can be introduced, and the Hamilton equations are
\begin{equation}
\partial_0\Pi(f)=\{\mathcal{H};\Pi(f)\}=\int d^3x fH=H(f)
\end{equation}
\begin{equation} 
\partial_0\vec{\Pi}(\vec{f})=\{\mathcal{H};\vec{\Pi}(\vec{f})\}=\int d^3x f_iH^i=\vec{H}(\vec{f}).
\end{equation}
From these equations one recognizes that the conservations of primary constraints implies the vanishing of the super-momentum and the super-Hamiltonian as secondary constraints, {\it i.e.}
\begin{equation}
H=0,\qquad H^i=0.\label{secconstr}
\end{equation}
These additional conditions are equivalent with non-evolutionary Einstein equations $G_{\mu0}=0$; this can be probed observing that
\begin{equation}
G_{\mu\nu}\eta^\mu\eta^\nu=-\frac{H}{2\sqrt{h}}\qquad G_{\mu\nu}e^\mu_ie^\nu_i=\frac{H_i}{2\sqrt{h}}.
\end{equation}
The algebra of constraints reads 
\begin{eqnarray} 
\{\vec{H}(\vec{f});\vec{H}(\vec{f'})\}=\vec{H}({\cal L}_{\vec{f}} \vec{f'})\label{alcons1}\\
\{\vec{H}(\vec{f});H(f')\}=\vec{H}({\cal L}_{\vec{f}}f')\label{alcons2}\\
\{H(f);H(f')\}=H(\vec{N}(f;f';h)),\label{alcons3} 
\end{eqnarray}
${\cal L}$ being the Lie derivative, while the components of $\vec{N}(f;f';h)$ are given by
\begin{eqnarray}
N^i(f;f';h)=h^{ij}(f\partial_jf'-f'\partial_jf).
\end{eqnarray}
From these relations it can be inferred that the constraints are first-class constraints (Poisson brackets of them are linear combinations of the same), thus no modification of the symplectic structure is required. A deep complication in view of the quantization is provided by the fact that the algebra of constraints is not a Lie one, since now one has to deal with structure functions and not structure constants anymore.\\  

\section{The Holst formulation}\label{holst}
The Holst reformulation of GR \cite{Ho96} consists in performing a first-order formulation and in adding a topological term to the action, {\it i.e.} a piece that vanishes as soon as equations of motion stand. The full action reads
\begin{eqnarray}
S_{G}=-\frac{c^{3}}{16\pi G}\int d^{4}x e e_\alpha^\mu e^\nu_\beta\bigg(R^{\alpha\beta}_{\mu\nu}-\frac{1}{2\gamma}\epsilon^{\alpha\beta}_{\phantom1\phantom2\gamma\delta} R^{\gamma\delta}_{\mu\nu}\bigg) 
\end{eqnarray} 
$\gamma$ being a free parameter (the Immirzi parameter). Variations with respect to connections $\omega^{\alpha\beta}_\mu$ provide
\begin{eqnarray}
\delta S=-\frac{c^3}{16\pi G}\int d^4xee^\mu_\alpha e^\nu_\beta\bigg(\delta R^{\alpha\beta}_{\mu\nu}-\frac{1}{2\gamma}\epsilon^{\alpha\beta}_{\phantom1\phantom2\gamma\delta} \delta R^{\gamma\delta}_{\mu\nu}\bigg)=\nonumber\\=\frac{c^3}{16\pi G}\int d^4x\delta {}^{(\gamma)}\!A_\nu^{\alpha\beta}\textit{D}_\mu(ee^\mu_\alpha e^\nu_\beta).\qquad\qquad\qquad
\end{eqnarray} 
The derivatives $\textit{D}_\mu$ act on space-time and on Lorentz indexes as follows
\begin{equation}
\textit{D}_\mu e^\nu_\alpha=\nabla_\mu e^\nu_\alpha-\omega_{\mu\alpha}^{\phantom1\phantom2\beta}e_\beta^\nu,
\end{equation}
while the connections ${}^{(\gamma)}\!A_\mu^{\alpha\beta}$ (Barbero-Immirzi connections \cite{B95}) take the form
\begin{equation}
{}^{(\gamma)}\!A_\mu^{\alpha\beta}=\omega^{\alpha\beta}_{\mu}-\frac{1}{2\gamma}\epsilon^{\alpha\beta}_{\phantom1\phantom2\gamma\delta}\omega^{\gamma\delta}_{\mu}.
\end{equation}
From the relation above, $\omega_\mu^{\alpha\beta}$ can be evaluated for $\gamma\neq\pm i$, having $\omega_\mu^{\alpha\beta}=\frac{\gamma^2-1}{\gamma^2}\bigg({}^{(\gamma)}\!A^{\alpha\beta}_{\mu}+\frac{1}{2\gamma}\epsilon^{\alpha\beta}_{\phantom1\phantom2\gamma\delta}{}^{(\gamma)}\!A^{\gamma\delta}_{\mu}\bigg)$. This demonstrates that arbitrary variations of $\omega_\mu^{\alpha\beta}$ provide arbitrary variation of $A_\mu^{\alpha\beta}$, hence the variational principle implies 
\begin{equation}
\textit{D}_\mu(ee^\mu_\alpha e^\nu_\beta)=0.\label{hstreq}
\end{equation}
This equation is the first Cartan structure equation for vanishing torsion. The second-order formulation is recovered by substituting the expression for $\omega_\mu^{\alpha\beta}$ coming out from (\ref{hstreq}). The additional Holst term into the action does not modify equations of motion: in fact, because of the cyclic identity for the Riemann tensor, it vanishes on-shell, {\it i.e.} 
\begin{equation}
\epsilon^{\alpha\beta}_{\phantom1\phantom2\gamma\delta} e e_\alpha^\mu e^\nu_\beta R^{\gamma\delta}_{\mu\nu}=\epsilon^{\mu\nu\rho\sigma}R_{\mu\nu\rho\sigma}=0.
\end{equation} 
Hence the Einstein and the Holst formulations are equivalent on the classical level.\\
For $\gamma=i$ and $\gamma=-i$ (Ashtekar connections \cite{As87}), ${}^{(\gamma)}\!A_\mu^{\alpha\beta}$ are the self-dual and the anti-self-dual part of Lorentz connections $\omega^{\alpha\beta}_\mu$, respectively, {\it i.e.} they satisfy \begin{equation}
\pm\frac{i}{2}\epsilon^{\alpha\beta}_{\phantom1\phantom2\gamma\delta}{}^{(\pm i)}\!A^{\gamma\delta}_{\mu}={}^{(\pm i)}\!A^{\alpha\beta}.
\end{equation}
This way, the 3 complex quantities ${}^{(\pm i)}\!A^{0a}_\mu$ determine the full connections $\omega^{\alpha\beta}_\mu$ in this case.\\ Into the expression $R^{\alpha\beta}_{\mu\nu}\pm\frac{i}{2}\epsilon^{\alpha\beta}_{\phantom1\phantom2\gamma\delta} R^{\gamma\delta}_{\mu\nu}$ only appears ${}^{(\pm i)}\!A^{\alpha\beta}_\mu$, and ${}^{(\mp i)}\!A^{\alpha\beta}_\mu$ does not. This means that variations are to be performed with respect to ${}^{(\pm i)}\!A^{\alpha\beta}_\mu$ and the first Cartan structure equation arises.\\
A very impressive way to rewrite the action, in terms of ${}^{(\pm i)}\!A^{0a}_\mu$, is the following one
\begin{equation}
S=-\frac{c^3}{16\pi G}\int d^4x e \left(e_0^\mu e_a^\nu\pm\frac{i}{2}\epsilon_a^{\phantom1 bc}e_b^\mu e_c^\nu\right){}^{(\pm i)}\!F^a_{\mu\nu}
\end{equation}
where ${}^{(\pm i)}\!F^a_{\mu\nu}=\partial_{[\mu}{}^{(\pm i)}\!A^{0a}_{\nu]}\pm \frac{i}{2}\epsilon^a_{\phantom1bc}{}^{(\pm i)}\!A^{0b}_{[\mu}{}^{(\pm i)}\!A^{0c}_{\nu]}$ is the curvature associated with SU(2) connections. The possibility to avoid the presence of others connections, in the case $\gamma=\pm i$, is a consequence of the isomorphism between the Lorentz group and the group $SU(2)\otimes SU(2)$, developed as the direct product of the self-dual and anti-self dual parts of $SO(1,3)$.\\ 
In the case $\gamma\neq\pm i$, this decomposition cannot be performed and the full action turns out to contain also ${}^{(-\gamma)}\!A^{\alpha\beta}_\mu$, even though it has no evolutionary character.\\ 
These speculations stress the peculiar role of Ashtekar connections.\\   

\paragraph{3+1 splitting} 
The 3+1 splitting procedure of the Holst action is based on taking vier-bein like quantities as configuration variables. This choice will allow to obtain additional constraints, a part coming out from the super-Hamiltonian and the super-momentun ones, which will be very useful when applying techniques proper of gauge theories to the quantization of gravity. \\ 
The following vier-bein is usually introduced 
\begin{equation}
e_\mu^{\phantom1\alpha}=\left(\begin{array}{cc} \widetilde{N} & \widetilde{N}^ie_i^a \\ 0 & e^a_i \end{array}\right)\label{3+1bein}
\end{equation}
{\it i.e.} the $e^0$ vector is fixed normal to $\Sigma$ (time gauge) and boost transformations have been frozen out. An explanation for this choice has been given by Barros e Sa \cite{BS01}, who demonstrated that, as soon as a generic vier-bein is taken into account, some second-class constraints arise. Solving them, he obtained just a first-class set, which he expected to be safety fixed before the quantization, by means of standard tools of gauge theories.\\  
Hence, within this scheme, the action rewrites as follows
\begin{eqnarray*}
S=-\frac{c^4}{16\pi G}\int dtd^3x\bigg [2E_a^i\bigg(\frac{1}{2}\partial_t{}^{(\gamma)}\!A^a_i-\frac{1}{2}(\partial_i\omega_t^{0a}-\frac{1}{2\gamma}\epsilon^{a}_{bc}\nonumber\\\partial_i\omega^{bc}_t)+\omega^{ab}_{[t}\omega_{i]b}^a-\frac{1}{2\gamma}\epsilon^{a}_{bc}\omega^{b0}_{[t}\omega_{i]0}^c-\frac{1}{2\gamma}\epsilon^a_{bc}\omega^{bd}_{[t}\omega_{i]d}^c\bigg)-\nonumber\\-2N^iE^j_a\bigg(R^{0a}_{ij}-\frac{1}{2\gamma}\epsilon^{a}_{bc}R^{bc}_{ij}\bigg)+ee^i_ae^j_b\bigg(R^{ab}_{ij}-\frac{1}{\gamma}\epsilon^{ab}_{\phantom1c}R^{c0}_{ij}\bigg)\bigg]
\end{eqnarray*}
with ${}^{(\gamma)}\!A^a_i={}^{(\gamma)}\!A^{0a}_i$ and $E^i_a$ densitized 3-bein $ee^i_a$.\\Hence, from the expression above, ${}^{(\gamma)}\!A^a_i$ are the only variables with an evolutionary character, while $\omega^{ab}_t$, $\omega_t^{0a}$ and ${}^{(-\gamma)}\!A^a_i$ behave as Lagrangian multipliers. By eliminating them from the action and after some algebra, one finally obtains 
\begin{equation}
S=-\frac{c^4}{16\pi G}\int dt d^3x (-\partial_t{}^{(\gamma)}\!A^a_i E^i_a+\frac{1}{\gamma}\Lambda^aG_a+N^iH_i+NH)
\end{equation}
$\Lambda^a$ being equal to $\frac{1}{2}\epsilon^a_{\phantom1bc}{}^{(\gamma)}\!A^{bc}_0-\frac{1}{\gamma}{}^{(\gamma)}\!A_0^{0k}$, while the new constraint reads
\begin{eqnarray}
G_a=\partial_iE^i_a+\gamma\epsilon^c_{ba}{}^{(\gamma)}\!A^b_iE^i_c\label{Gcon}\\H_i=E^j_a{}^{(\gamma)}\!F^a_{ij}\label{supmome}\\H=\frac{1}{\gamma}\epsilon_i^{\phantom1jk}e^i_a({}^{(\gamma)}\!F^a_{jk}-\frac{\gamma^2+1}{2\gamma}R_{jk}^{\phantom1\phantom2a}),\label{supHam}
\end{eqnarray}
with ${}^{(\gamma)}\!F^{a}_{\mu\nu}=\partial_{[i}{}^{(\gamma)}\!A^a_{j]}+\frac{\gamma}{2}\epsilon^a_{\phantom1bc}{}^{(\gamma)}\!A^b_{[i}{}^{(\gamma)}\!A^c_{j]}$, and $R_{ij}^a=\partial_{[i}\Gamma^a_{j]}+\epsilon^a_{\phantom1bc}\Gamma^b_i\Gamma^c_j$ ($\Gamma^a_i=-\frac{1}{2}\epsilon^a_{\phantom1bc}e^b_j\nabla_ie^{jc}$).\\One ends up with a phase space having connections ${}^{(\gamma)}\!A^a_i$ and densitized triads as conjugate variables, while the additional constraint is the Gauss constraint in a $SU(2)$ gauge theory.\\\\
For $\gamma=\pm i$, two main points have to discussed. The first one is that the time-gauge condition is not requested, being Ashtekar connections ${}^{(\pm i)}\!A^a_{i}$ the pull-back on $\Sigma$ of space-time ones, {\it i.e.} ${}^{(\pm i)}\!A^{ab}_\mu$. This point is related with the second one: because of the clear geometrical interpretation of Ashtekar connections, the super-Hamiltonian constraint, which generates time-reparametrizations, takes a much simpler form: in fact, it turns out that, except for a square root of the 3-metric, the super-Hamiltonian, and thus the full set of constraints, is polynomial in the configuration variables. This feature will provide a significantly simplification in the quantum context.\\ The difficulties of a formulation in terms of Ashtekar connections come from their complex character. This implies that the $SU(2)$ group one is dealing with is a non-compact group, for which techniques proper of gauge theories cannot be applied. For instance, one is not able to implement reality conditions \cite{Ro91} on triads in a quantum context. Therefore, real connections are preferred with respect to Ashtekar connections, despite the clear geometrical meaning of the latter.\\

\section{Boost invariance in Loop Quantum Gravity}
The machinery of Loop Quantum Gravity is based on performing a quantization in terms of a non-canonical algebra, the holonomy-flux one. The request of 3-diffeomorphism invariance singles out a unique representation \cite{AL95}, such that a Hilbert space can be properly defined. One of the main achievements of this formulation is the evaluation of geometric operators' spectra (areas and volumes), which turns out to be discrete. What seems not to be clear is the behavior of these spectra under boosts, {\it i.e.} if, as for macroscopic geometric objects, they are subject to some sort of Lorentz contraction.\\ 
The work by Barros e Sa \cite{BS01} suggests that if these geometrical operators are observables, then their spectra cannot be modified by a boost, this one being a gauge transformation. The work by Rovelli and Speziale also points in this direction \cite{RS06}, in which a comparison with the behavior of the angular momentum operators in ordinary quantum mechanics is made. However Alexandrov \cite{Al00} performed a covariant formulation, in which no gauge fixing has been provided and second-class constraints were treated replacing Poisson brackets with Dirac ones. A Gauss constraint for the Lorentz group was identified and, although the Hilbert space could not be defined (because configurations variables do not commute), nevertheless the spectra of some geometric operators were found. These spectra contain a dependence on the Lorentz frame.\\ 
Therefore, we are going to perform a formulation free of the time gauge, with the aim of realizing a quantum transformation between different frames, such that the invariant or non-invariant character of operators' spectra can be verified.

\section{Hamiltonian formulation without the time gauge}
In this section we are going to provide an Hamiltoian formulation with configuration variables describing an arbitrary vier-bein in the space-time, which can be written as
\begin{equation}
e_\mu^{\phantom1\alpha}=\left(\begin{array}{cc} N & N^iE_i^a \\ \chi_a E^a_i & E^a_i \end{array}\right).\label{fram}
\end{equation}
The relations between $N$, $N^i$, $\chi_a$ and $E^a_i$ and quantities proper of the ADM splitting are the following ones
\begin{eqnarray}
\left\{\begin{array}{c}
\tilde{N}=\frac{1}{\sqrt{1-\chi^2}}(N-N^iE_i^a\chi_a) \\\\ 
 \tilde{N}^i=N^i+\frac{E^c_l\chi_cN^l-N}{1-\chi^2}E^i_a\chi^a\\\\ 
h_{ij}=E^a_iE^b_j(\delta_{ab}-\chi_a\chi_b)
\end{array}\right.
\end{eqnarray}
and they outline how, for $\chi_a=0$ the time-gauge condition is restored. A way to work out such condition consists in performing a Lorentz transformation $\Lambda^A_{\phantom1B}$, with $\chi_a=-\Lambda^0_{\phantom1a}/\Lambda^0_{\phantom10}$. Hence $\chi_a$ is the velocity components of the frame (\ref{fram}) with respect to spatial hypersurfaces.\\ 
Starting from the Lagrangian of the action (\ref{azsplit}), momenta associated with the set of configuration variables $\{\widetilde{N},\widetilde{N}^i,E^a_i,\chi_a\}$ are given by
\begin{eqnarray}
\pi_{\tilde{N}}=0\qquad\pi_i=0\\
\pi^i_a=\frac{1}{8\pi G}\sqrt{h}\left[K^{ij}E_j^b(\delta_{ab}-\chi_a\chi_b)-KE_a^i\right]\\
\pi^a=\frac{1}{8\pi G}\sqrt{h}\bigg(\frac{\chi^a}{1-\chi^2}K-K^{ij}E^a_iE^b_j\chi_b\bigg).
\end{eqnarray}
From these relations, phase space variables turn out to satisfy the following conditions
\begin{eqnarray}
\pi_{\tilde{N}}=0\qquad\pi_i=0\label{piNNi}\\
\Phi^a=\pi^a-\pi^b\chi_b\chi^a+\delta^{ab}\pi^i_b\chi_cE^c_i=0\label{boost}\\
\Phi_{ab}=\pi^c\delta_{c[a}\chi_{b]}-\delta_{c[a}\pi^i_{b]}E^c_i=0.\label{rot}
\end{eqnarray}
This way, the full Hamiltonian can be written as
\begin{equation}
\mathcal{H}=\tilde{N}'H+\tilde{N}^iH_i+\lambda^{\tilde{N}}\pi_{\tilde{N}}+\lambda^i\pi_i+\lambda^{ab}\Phi_{ab}+\lambda_a\Phi^a.
\end{equation}
$\lambda^{\tilde{N}}$, $\lambda^i$, $\lambda_a$ and $\lambda^{ab}=-\lambda^{ba}$ being Lagrangian multipliers which keep constraints on the dynamics. The new super-Hamiltonian and super-momentum read as follows 
\begin{eqnarray}
H=\pi^i_a\pi^j_b\bigg(\frac{1}{2}E^a_iE^b_j-E^b_iE^a_j\bigg)+h{}^{3}\!R\label{H}\\
H_i=D_j(\pi^j_aE^a_i),\label{Hi}
\end{eqnarray}
and their vanishing behavior comes again as secondary conditions from relations (\ref{piNNi}).\\ 

\section{Properties of constraints}
Constraints acquire a precise physical meaning as soon as their action on the phase-space is investigated. We quote the well-know result that the four standard conditions (\ref{piNNi}), together with the secondary ones, {\it i.e.} the vanishing behavior of the super-Hamiltonian and of the super-momentum, accounts for the invariance under time re-parametrization and spatial diffeomorphisms, respectively.\\
Other constraints enforce the invariance under 4-bein Lorentz transformations: in fact we have 
\begin{eqnarray}
\{\Phi^a;e^0\}=E^a_idx^i\\\{\Phi^a;e^c\}=\delta^{ac}E^d_i\chi_dN^idt+\delta^{ac}E^d_i\chi_ddx^i\\
\{\Phi_{ab};e^0\}=0\qquad\{\Phi_{ab};e^c\}=\delta_{d[a}\delta^c_{b]}e^d,
\end{eqnarray}
which outline how $\Phi_{ab}$ generates a rotation of the vier-bein, while the transformation associated with $\Phi^a$ is a combination of a boost and a time re-parametrization.\\   
As a consequence of this scheme, once $\varphi^a=\epsilon^{abc}\Phi_{bc}$ is introduced, the boost-rotation algebra is reproduced
\begin{eqnarray}
\{\Phi^a;\Phi^b\}=\epsilon^{ab}_{\phantom1\phantom2c}\varphi^c\quad\{\varphi^a;\varphi^b\}=-\epsilon^{ab}_{\phantom1\phantom2c}\varphi^c\\\{\varphi^a;\Phi^b\}=-\epsilon^{ab}_{\phantom1\phantom2c}\Phi^c.\label{boost-rot}
\end{eqnarray} 
The Dirac algebra of constraints is a fundamental feature in view of the quantization, since it gives the classification in first-class and second-class sets. The latter are associated with dynamical systems in which constraints can be eliminated by a suitable choice of variables, while the former describe a scenario with gauge symmetries. Here, the additional constraint $\varphi_a=0$ may change the set of constraints from being first-class to second-class.\\ 
The first-class character of $\{\pi,\pi_i,H,H_i,\phi_{ab}\}$ is a well-know result from quantization within the time gauge. In presence of the condition $\Phi^a=0$, the transformations generated by $\Phi^a$ have to be investigated. As far as the action on the 3-metric is concerned one finds 
\begin{equation}
\{\Phi^a;h_{ij}\}=0.
\end{equation} 
Furthermore, Poisson brackets of the terms appearing into the super-Hamiltonian and super-momentum give
\begin{eqnarray}
\{\Phi^a;\pi_c^iE_j^c\}=0\\\{\Phi^a;\pi^i_c\pi^j_d\bigg(\frac{1}{2}E^c_iE^d_j-E^d_iE^c_j\bigg)\}=0,
\end{eqnarray}
which, together with the previous one, allows to conclude that 
\begin{equation}
\{\Phi^a;H\}=\{\Phi_{ab};H\}=0\qquad\{\Phi^a;H_i\}=\{\Phi_{ab};H_i\}=0.\label{br-hm}
\end{equation}
Therefore, the first-class property of the full set of constraints is not modified by the introduction of the boost.

\section{Gauge fixing and Quantization}
Let us now fix, formally, the boost symmetry, before the quantization. One can set $\chi_a$ equal to some space-time functions $\bar{\chi}_a(t;x)$, such that they are no more phase-space variables. Conjugated momenta can be expressed by solving $\Phi^a=0$, so having   
\begin{equation}
\pi^a=-\bigg(\delta^{ab}+\frac{\bar{\chi}^a\bar{\chi}^b}{1-\bar{\chi}^2}\bigg)\pi^i_b\bar{\chi}_cE^c_i.
\end{equation}
This way, the gauge-fixed action reads
\begin{eqnarray}
S=\int[\pi^i_a\partial_tE^a_i+\pi_{\tilde{N}'}\partial_t\tilde{N}'+\pi_i\partial_t\tilde{N}^i-\tilde{N}'H^{\bar\chi}-\nonumber\\-\tilde{N}^iH^{\bar\chi}_i-\lambda^{ab}\Phi'_{ab}-\lambda^{\tilde{N}}\pi_{\tilde{N}}-\lambda^i\pi_i]dtd^3x,
\end{eqnarray}
where a new rotation constraint arises, having the following expression
\begin{equation}
\Phi'_{ab}=\bar{\chi}_{[a}\pi^i_{b]}E^d_i\bar{\chi}_d-\delta_{c[a}\pi^i_{b]}E^c_i,\label{rotcon}
\end{equation}
while into the super-Hamiltonian and the super-momentum ${\bar\chi}_a$ replace $\chi_a$.\\
This formulation clearly indicates that the dynamics depends also on $\bar{\chi}$, {\it i.e.} on the Lorentz frame. This feature is related to the geometrical interpretation of $E^a_i$, which is the projection of the 3-bein on spatial hypersurface, hence their values depend on $\bar{\chi}_a$ (which gives the projection on the time axis). There is no way to avoid such a dependence, neither by taking them as coordinates.\\
The canonical quantization procedure consists in taking  $\tilde{N}$, $\tilde{N}^i$ and $E^a_i$ as multiplicative operators and in replacing Poisson brackets times $i\hbar$ with commutators. Hence a representation for operators corresponding to $\pi_{\tilde{N}}$, $\pi_i$ and $\pi_a^i$ is given by $i\hbar\frac{\delta}{\delta\tilde{N}}$, $i\hbar\frac{\delta}{\delta\tilde{N}^i}$ and $i\hbar\frac{\delta}{\delta E^a_i}$, respectively.\\ 
According with the Dirac prescription, constraints are translated into conditions physical states must satisfy. As far as conditions (\ref{piNNi}) are concerned,  they imply $\frac{\delta}{\delta\tilde{N}}\psi=\frac{\delta}{\delta \tilde{N}^i}\psi=0$, thus, into the full Hilbert space, physical states $\psi$ belong to the subspace of functional independent on $\tilde{N}$ and $\tilde{N}^i$. Furthermore the super-momentum constraint takes the following expression
\begin{equation}
\hat{H}^{\bar\chi}_i\psi_{\bar\chi}(E)=iD_j\bigg(E^a_i\frac{\delta}{\delta E^a_j}\bigg)\psi_{\bar\chi}(E)=0\label{qsm}
\end{equation}
and it gives that wave functionals are unchanged by $E^a_i\rightarrow E^a_i-D_i\xi^jE^a_j$, being $\xi^i$ an arbitrary 3-vector. This is the way the 3-diffeomorphism invariance is implemented on a quantum level.\\ 
Then the rotational constraint becomes
\begin{equation}
\Phi'^{\bar\chi}_{ab}\psi_{\bar\chi}(E)=i\bigg[{\bar\chi}_{[a}\frac{\delta}{\delta E_i^{b]}}E^d_i{\bar\chi}_d-\delta_{c[a}\frac{\delta}{\delta E_i^{b]}}E^c_i\bigg]\psi_{\bar\chi}(E)=0:\label{qrotcon}  
\end{equation}
and these conditions emphasize how a parametric dependence on $\bar{\chi}$ cannot be avoided.\\ 
Finally, the super-Hamiltonian constraint provides the dynamics and it reads 
\begin{equation}
\hat{H}^{\bar\chi}\psi_{\bar\chi}(E)=\bigg[-\bigg(\frac{1}{2}E^a_iE^b_j-E^b_iE^a_j\bigg)\frac{\delta}{\delta E_i^a}\frac{\delta}{\delta E_j^b}+h{}^{3}\!R\bigg]\psi_{\bar\chi}(E)=0.\label{qsh}
\end{equation}

\section{Quantum boosts}
The formulation of the previous section can be applied to a generic Lorentz frame, as soon as the form of functions $\bar{\chi}_a$ is given. But the main point of this treatment is the possibility to implement transformations between Hilbert spaces with different $\bar{\chi}$, hence it allow the investigation on the fate of the boost invariance.\\ 
In this scheme, quantum boost can be represented starting from transformations generated by $\Phi^a$ on the phase space and developing the quantum analogous. In particular, the following expression arises on the hypersurface $\chi_a=0$  
\begin{equation}
U_\epsilon=I-\frac{i}{4}\int\epsilon^a\epsilon_b(E^b_i\pi^i_a+\pi^i_aE^b_i)d^3x+O(\epsilon^4),\label{U}
\end{equation}
$\epsilon_a$ being arbitrary infinitesimal parameters.\\ 
In order to recognize a transformation between Hilbert spaces as a symmetry, scalar products must not be modified and physical states on the first Hilbert space have to be sent into physical states on the second one. This means that the transformation (\ref{U}) must be unitary and that it must map solutions of constraints from the $\bar{\chi}_a=0$ to the $\bar{\chi}_a=\epsilon_a$ sectors.\\ 
While the unitary character of the operator (\ref{U}) is manifest, being generators Hermitian, the second request has to be explicitly checked.\\  
A physical state for $\bar{\chi}_a=0$ satisfies the following conditions
\begin{equation}
H^{0}\psi_0=0\quad H^{0}_i\psi_0=0\quad-\delta_{c[a}\pi^i_{b]}E^c_i\psi_0=0,
\end{equation}
with $H^{0}$ and $H^{0}_i$ the super-Hamiltonian and super-momentum for $\bar{\chi}\equiv0$.\\
Under the action of $U$, the new state $\psi'(E)=\psi_0(E')$ satisfies
\begin{eqnarray}
U_\epsilon H^0U_\epsilon^{-1}\psi'=0\qquad U_\epsilon H^0_iU_\epsilon^{-1}\psi'=0\\ U_\epsilon(-\delta_{c[a}\pi^i_{b]}E^c_i)U^{-1}_\epsilon\psi'=0.
\end{eqnarray}
Starting from the following conditions
\begin{eqnarray} 
U_\epsilon\delta_{ab}E^a_iE^b_jU_\epsilon^{-1}=E^a_iE^b_j(\delta_{ab}-\epsilon_a\epsilon_b)+O(\epsilon^4)\\
U_\epsilon
E^a_i\pi^j_aU_\epsilon^{-1}=E^a_i\pi^j_a+O(\epsilon^4),
\end{eqnarray}
one recognizes that $U_\epsilon$ maps $H^{0}$ and $H^{0}_i$ into $H^\epsilon$ and $H^\epsilon_i$, respectively, up to the $\epsilon^2$ order.\\ 
As far as the rotation constraint is concerned, the condition obtained on $\psi'$ is as follows
\begin{eqnarray}
-\bigg[\delta_{c[a}\pi^i_{b]}E^c_i+\frac{1}{2}\delta_{c[a}\epsilon_{b]}\epsilon^dE^c_i\pi^i_d-\nonumber\\\frac{1}{2}\epsilon_d\epsilon_{[a}\pi^i_{b]}E^d_i\chi_d+O(\epsilon^4)\bigg]\psi'=0.\label{Ucon}
\end{eqnarray}
From the relation above, the action of $\epsilon_dE^d_i\pi^i_b$ on $\psi'$ can be evaluated, by virtue of a multiplication times $\epsilon^a$ and by retaining the leading orders in $\epsilon_a$. If the expression obtained is substituted into the condition (\ref{Ucon}), the vanishing of the rotation constraint in the $\bar{\chi}_a=\epsilon_a$ arises, {\it i.e.}
\begin{equation}
\bigg[\delta_{c[a}\pi^i_{b]}E^c_i-\epsilon_d\epsilon_{[a}\pi^i_{b]}E^d_i\chi_d+O(\epsilon^4)\bigg]\psi'=0.
\end{equation}  
These results indicate that the boost operator $U_\epsilon$ realizes a symmetry between Hilbert spaces of the $\bar{\chi}_a=0$ and $\bar{\chi}_a=\epsilon_a$. Therefore the boost invariance is not affected by a canonical quantization of gravity, in a second-order formulation.

\section{Concluding remarks}
The canonical quantization of General Relativity is usually performed within the time gauge, {\it i.e.} the Lorentz frame is fixed with the time-like vector orthogonal to spatial hypersurfaces. This feature provides a quantum description for the 3-geometry only in a frame co-moving with spatial hypersurfaces. By dropping this assumption, some additional variables, describing the velocity component of the frame, are present, together with 3 new constraints, related to the boost invariance. Such a symmetry can be fixed before the quantization and no quantum violation arises. This result has been probed by the direct development of an operator corresponding to infinitesimal boosts on the Hilbert space. Hence, this operator is unitary and realizes the mapping of the set of physical states into itself.\\ 
In this formulation, one concludes that the spectrum of observables must be the same in any Lorentz frame. Prospectives of this work deal with a much more realistic description with matter fields, which give a 4-diffeomorphism invariant meaning to spatial hypersurfaces. Moreover, the development of a first-order formulation, in terms of Ashtekar-Barbero-Immirzi connections, is a first step towards the extension of this framework to Loop Quantum Gravity. In these models discrete spectra are predicted for area and volume operators and the determination of their behavior under boosts is a great result, in view of characterizing a quantum space-time. In fact, this point can give a connection with phenomenological approaches to Quantum Gravity, like some non-commutative models \cite{AC02}.

\end{document}